\documentclass{article}   
\usepackage{natbib}

\usepackage{physics,amsmath,amssymb,mathrsfs}
\usepackage{enumitem}

\title{Trans-Planckian Philosophy of Cosmology}
\author{Mike D. Schneider\footnote{I would like to thank Jeremy Steeger and Adam Koberinski for their early reactions to this project, and I would also like to thank my (virtual) audience at the Beyond Spacetime group in November 2020. More recently, I am indebted to two very helpful anonymous reviewers. This project was conceived after my attending the Conference on the Foundations of Cosmology and Quantum Gravity at NYU Abu Dhabi in January 2020. There, on the first day, Robert Brandenberger introduced the trans-Planckian censorship conjecture as a still quite new and controversial proposal in theoretical physics. That talk, along with various ensuing discussions in the following days, convinced me that there is philosophical work to be done alongside the new physics, even at the earliest stages. I hope that the present article will be seen in this light.}}
\date{}

\begin{document}
\maketitle

\begin{abstract}
I provide some philosophical groundwork for the recently proposed `trans-Planckian censorship' conjecture in theoretical physics. In particular, I argue that structure formation in early universe cosmology is, at least as we typically understand it, autonomous with regards to quantum gravity, the high energy physics that governs the Planck regime in our universe. Trans-Planckian censorship is then seen as a means of rendering this autonomy an empirical constraint within ongoing quantum gravity research.
\end{abstract}


\section{A new conjecture}\label{secnottealeaves}

In September 2019, two papers were posted to the physics preprint repository ArXiv, which together introduced the trans-Planckian Censorship Conjecture (TCC) to the community engaged at the intersection of theoretical physics and early universe cosmology \citep{bedroya2019trans, bedroya2020trans}. Although the TCC was originally motivated as a particular swampland conjecture in string theory (more on this in section \ref{secTCC}), the relationship between the two subjects might now be inverting (e.g. \citep{bedroya2021sitter})--- even while controversy over the original proposal in string theory rages on (e.g. \citep{saito2020trans}). As a result, `trans-Planckian censorship' seems poised to become a subject of semiclassical study in its own right at the frontiers of quantum gravity research, semi-independent of questions concerning its formulation as any one conjecture (i.e. as would be relevant in any one candidate quantum gravity approach, like string theory). 

All for the better! As I will here argue, trans-Planckian censorship, so generalized, serves well to couple together one typical way of thinking about how to model structure formation in the early universe (as is meanwhile constrained by cosmological data) with our ongoing theorizing about quantum gravity. Still, its original connection to the swampland conjectures in string theory--- technical conjectures about which effective field theories fail to admit stringy high-energy completions--- remains significant. My task below will be, in part, to articulate what persists of that original connection.

Here is the TCC, in slogan form:
\begin{quote}Nature precludes the physical relevance of a large class of semiclassical field theories defined on cosmological spacetimes. \end{quote}
The specific terms in this slogan will be discussed in the following section. Roughly speaking, trans-Planckian censorship stipulates that the high energy physics of the Planck regime is sequestered, necessarily, from certain typical descriptions we give of structure formation in our remote past. In other words, this practically rules out the viability that certain quantum fluctuations, which would otherwise encode unknown quantum gravitational physics, ultimately seed the evolution of large scale structure observed today \citep{bedroya2019trans}.\footnote{There is also a question about the role of the TCC in the asymptotic future of the universe (e.g., \citep{heisenberg2021model}), in relation to the cosmological constant problem. But such an application of the TCC would seem to require, in its theoretical background, our taking seriously a `strong' version of the trans-Planckian problem in cosmology (discussed below). As will become clear, there is at least one sense in which taking seriously a weaker version of the trans-Planckian problem, which is relevant to the application of the TCC in early universe cosmology, requires our simultaneously setting aside the strong version. And since the popular application of the TCC concerns the early universe, I take this to be sufficient cause as to set aside, for now, the further topic noted presently.}

More on the details, shortly. For now, one might criticize of my slogan that the language ``nature precludes'' is obscure. But this is intentional: it emphasizes the peculiar status of the subject of trans-Planckian censorship, given current physics. Namely, whereas facts about string theory might (by the original version of the TCC) entail the conclusion just rehearsed (cf. section \ref{secTCC}), the epistemic warrant for that conclusion, in such a case, is conditional on string theory coming to provide an adequate theory of quantum gravity. This is so, \emph{even if we grant that the relevant version of the conjecture is true}, i.e. that the swampland conjecture that would ensure trans-Planckian censorship in string theory is essentially correct. Meanwhile, there are, conceivably, other facts, conditional on the successes of other candidate quantum gravity approaches besides string theory, that could likewise entail the same conclusion, just by other means (i.e. by other versions of the TCC, which we may likewise grant are true). In this sense, the most general, disjunctive form of the TCC--- as summarized in the slogan above--- will clearly remain open for \emph{at least} as long as we lack an adequate theory of quantum gravity. Only having achieved that sought-after theory will we subsequently be able to assign (by conjecture) a truth-value to trans-Planckian censorship: either that theory supplies such facts as would (by means of one particular version of the TCC) be appropriate to secure the conclusion, or else it does not.\footnote{\label{fnnatureprecludes}This echoes the observation I have made in \citep{schneider2020s} about the status of the cosmological constant problem, prior to achieving a future theory of quantum gravity. It also echoes the textbook discussion about cosmic censorship found in \citep[p. 304]{wald1984}, at least provided that the physical reasonableness of a relativistic spacetime is specifically understood as ultimately grounded in facts about quantum gravity (i.e. not just general relativity). These parallels suggest that there is nothing new, methodologically speaking, about something like trans-Planckian censorship coming to acquire a life of its own at the semiclassical frontiers of quantum gravity research, and to thereby shape ongoing research in the discipline. Moreover, at least in the case of the cosmological constant problem, I have argued in \citep{schneiderbetting} that this state of affairs can be epistemically well founded.}

As just presented, trans-Planckian censorship concerns a novel understanding of the relationship between lines of evidence in cosmology and theory development in quantum gravity. In fact, my principle contribution in the following is to provide a perspective on the relationship between lines of evidence in cosmology and theory development in \emph{high energy particle physics}, which is intended to be faithful to how \emph{that} relationship is typically understood, in practice. But, crucially, from this perspective, trans-Planckian censorship is rendered productive for furthering frontier physics research, when one turns to consider the prospects of an analogous relationship between cosmology and even higher energy physics, i.e. quantum gravity. Hence, I claim that this pragmatic upshot--- given the accuracy of the perspective just sketched--- is what justifies that trans-Planckian censorship is poised to take on a life of its own at the semiclassical frontiers of quantum gravity research. Contrary to standard presentations of the subject, the TCC would seem to do real, substantive work, for \emph{any} quantum gravity theorist who endeavors to wield it.\footnote{\label{fnworkinghypothesis} In a sense, the claim here is that a theorist's embrace of trans-Planckian censorship amounts to their adopting a working hypothesis, specifically within the context of quantum gravity research, about a lack of actual trans-Planckian physics within the early history of our universe (at least, as we typically understand that history, by means of positing an inflationary epoch). Note that this framing leaves open the possibility that our expectations about the future theory of quantum gravity could, presently, serve as impetus for revising our thinking about early universe structure formation in terms of inflationary dynamics. Nonetheless, here I focus on the other direction: where our holding fixed one typical way of thinking about cosmology empirically constrains (that is, by means of this working hypothesis) ongoing quantum gravity research.}

To the extent that the perspective I provide captures the kernel of the community's enthusiasm for trans-Planckian censorship, separate from its origins in string theory, the perspective ought to provide a suitable foundation for further philosophical engagement with the general subject (including potential disagreement with my pragmatic defense of its embrace). That being said, the elephant in the room is that trans-Planckian censorship is still new, and so it may seem premature to discuss its prospects. This is in sharp contrast with other similar cases of philosophical interest at the semiclassical frontiers of quantum gravity research (e.g. cosmic censorship or the cosmological constant problem--- cf. footnote \ref{fnnatureprecludes}), where the legacy of the relevant subject so far in history might appear to be the primary justification for philosophically studying it. 

In disanalogy with those other cases, one cannot presently appeal to any descriptive claim about the staying power of trans-Planckian censorship, as motivation for a philosophical investigation of the subject. Moreover, if the TCC is abandoned sufficiently quickly, it likely will not have left enough of a mark on the history of the discipline, for such a mark to merit philosophical scrutiny, in retrospect. And while I would like to think of myself as having an ear to the ground (so as to justify the project, provisionally, on the prediction that such doom and gloom will not be the case), it can be quite difficult to spell out the difference between an act of measured foresight and that of reading tealeaves. 

More preferable, then, is to locate a reason that motivates broad philosophical engagement with the new subject, which is not reducible to a prediction about the future winds of research in the discipline. In fact, intuitions along these lines might even pull in the opposite direction: perhaps the most responsible philosophical project concerning trans-Planckian censorship, at this nascent stage, is one which steers any such predictions, in the first place. So, for instance, \citet{dawid2013string} offers a framework for generating confirmation in the absence of novel empirical constraints. Along these lines, one might consider whether there is sufficient confirmation to license trust (cf. \citep{dawid2018delimiting}) in trans-Planckian censorship, and thereby serve as epistemic warrant for its uptake, right now. And if (as one might privately suspect) there is \emph{insufficient} confirmation to trust trans-Planckian censorship--- at least, unconditional on string theory, which is itself contentious--- one might regard the philosopher's greatest possible contribution to be the act of nipping the burgeoning general study of trans-Planckian censorship in the bud. Whereas the TCC might happen to live on (or not) as a particular swampland conjecture within string theory, its status therein would thereby ever remain an artifact of the technical details of that particular candidate quantum gravity approach. The further study of trans-Planckian censorship would simply not be pursued, separate from those technical details. 

There are, I believe, two responses to this suggestion. The first is procedural: to engage in such a philosophical project as could nip the study of trans-Planckian censorship in the bud, a certain amount of work is already necessary to interpret the subject, thereby disentangling its particulars from the particulars of intimately related others, e.g. string theory. For instance, as will be the focus of section \ref{secTCC}, the general subject requires certain formal infrastructure, in any given candidate quantum gravity approach, which is similar to that provided by the swampland in string theory. But, importantly, that infrastructure only \emph{needs to be} similar in one particular sense. As such, evidential arguments in support of trans-Planckian censorship may not be as constricted as might otherwise be thought.

This may be a sufficient response to motivate philosophical engagement of the kind I have in mind, and for which the perspective I provide in what follows can serve as a foundation. But the second response is more programmatic--- and thereby (I suspect) more compelling. Namely, insofar as wielding the TCC is consonant with the epistemic aims of the relevant community, independent of its formulation within, e.g., string theory, the pragmatic upshots of trans-Planckian censorship serve, \emph{in themselves}, as epistemic warrant for its uptake.\footnote{Though, it may be that its uptake is better thought about in terms other than belief--- cf. footnote \ref{fnendorsement} in the Conclusion.} Thus, to have put a finger on the pragmatism of a completely general rendition of trans-Planckian censorship is to simultaneously have located a reason for philosophical engagement with the new subject. Namely, it is the subject's \emph{genuine capacity to shape} future research in the discipline that makes it so interesting, and which justifies philosophers diving in, at so early a juncture, to discuss it. 

\section{Cosmology and high energy physics}\label{sectionCosmologyandHEP}

The aim of this section is to clarify the relationship between cosmology and high energy particle physics. As will become clear, this relationship is crucial for understanding how it is that structure formation in the early universe might empirically constrain quantum gravity research, by means of trans-Planckian censorship. 

First, some groundwork. The standard model of cosmology, $\Lambda$CDM, constitutes our current best theory of the evolution of large scale structure (LSS) in cosmology. As a consequence of our embrace of $\Lambda$CDM, we are committed to the descriptive accuracy of a particular narrative of large scale cosmic history. Namely, to zeroth approximation, our universe exhibits (always strictly positive) `cosmic expansion', wherein space has been uniformly expanding and cooling for all cosmic time. The evolution of any further, non-trivial LSS is then studied perturbatively, given cosmic expansion in the background.\footnote{That is, the study of non-trivial LSS is defined on `cosmological spacetimes': globally hyperbolic spacetimes foliated by spacelike hypersurfaces satisfying the cosmological principle. The leaves of this foliation are threaded by the integral curves of a unit timelike vector field that is everywhere hypersurface orthogonal (the spatial universe is assumed to be non-rotating), defining cosmic time. These integral curves describe the possible worldlines of stationary observers who bear witness to (possibly degenerate, or even negative) cosmic expansion. For more about these spacetimes and their physical interpretation, see \citep[ch. 2]{malament2012topics}.}  

Consequent to this narrative, as one travels backward through cosmic time about any point, LSS approximately uniformly contracts, growing hotter and hotter. Increasing temperatures, we eventually regard a semiclassical interpretation of quantum field theory on curved spacetime (QFTCS), familiar from high energy particle physics, as descriptively relevant to questions about the formation of any non-trivial LSS (still given cosmic expansion in the background). Sometimes, as in the case of inflationary cosmology (discussed below), these QFTCS constructions are further tasked with explaining, by means of a semiclassical treatment of gravity, cosmic expansion itself during that period of suitably high temperatures. In such a case, we may regard the construction as providing a model of structure formation relevant in our remote past \citep{azhar2017scientific}. The embrace of any one particular model of structure formation is thereby understood to modify what we take to be descriptively accurate of the zeroth-order approximation of the spatiotemporal universe during that early period. In this sense, the model of structure formation replaces the account otherwise given by $\Lambda$CDM, restricted to our remote past.

What I am discussing presently is the domain of `early universe cosmology', as the subject is understood in contemporary cosmological practice. (And so, models of structure formation are, equivalently, models of early universe cosmology.) In this domain, the semiclassical observables we care about most are various correlations produced along the surface of last scattering (relative to us) in a period of cosmic history following that of the `early universe', known as `recombination'. These correlations at recombination are what we may observe today, via measurements of the cosmic microwave background (CMB).

The state of affairs in early universe cosmology may be contrasted with that which follows it.\footnote{\label{fnsemiclassicalinterpolation}In fact, there is considerable freedom as to where in cosmic history before recombination we place the boundary of the domain of early universe cosmology. This is because the boundary of that domain arguably designates, in the first place, a transition before which one must attend to specifically quantum processes in modeling the evolutionary history of our universe in accordance with our local physics (due to thermal arguments, like those just sketched in the main text). But this view of the boundary renders the modeler's choice of where to draw it within our cosmic history dependent on the physical scales that are of interest to that particular modeler. For instance, at the scales relevant to the study of non-trivial LSS treated perturbatively--- as the large-scale universe has been presented here--- it is not obviously important that the universe remains, at much smaller scales, like a plasma up until recombination. That is, from the perspective of a modeler focused exclusively on the former, recombination is itself just not a particularly noteworthy transition in the evolutionary history of our universe; the modeler may apply the same dynamical equations continuously through the period. Still, since the semiclassical observables we typically care about most regarding this domain are features of the CMB, understood as traces of the physics present at recombination (when, thinking now about the much smaller scales relevant to our astronomical observations, the universe first became optically transparent), the period is often marked as a transition and the boundary drawn accordingly, even at the largest scales. Of course, this glosses over a lot of other interesting and dynamically relevant physics: one could also choose to draw the boundary much earlier, treating the ensuing gap in cosmic history between that boundary and recombination as a period that semiclassically interpolates between the two domains presently distinguished, respectively, as early universe and late-stage cosmology. I thank a reviewer for pressing me on this point.} Namely, in `late-stage cosmology', the same measurements of the CMB are taken to provide a means to infer suitable initial data at recombination for studying the evolution of LSS ever since. In this context, per the dictates of $\Lambda$CDM, a classical, relativistic perturbation theory is taken to be descriptively accurate of the dynamics of non-trivial LSS (given cosmic expansion in the background), instead of some QFTCS construction. This classical theory admits a well-posed Cauchy problem specified at recombination, and is assumed to approximate a fully self-consistent, general relativistic description of the evolution of LSS (i.e. including background) at all times following recombination. The upshot is that we may understand the high energy particle physics of early universe cosmology as \emph{seeding} the evolution of LSS in late-stage cosmology: structure formation in the former constrains the Cauchy problem specified at recombination in the latter. (And so, likewise, the Cauchy problem in the latter constrains structure formation in the former.)

There are a number of provisos one might wish to add to this presentation. The one that will be most important below concerns inflationary cosmology. Broadly speaking, `inflationary cosmology' refers to a class of scalar quantum field theory constructions on cosmological spacetimes, which are proposed as models of early universe cosmology. Given inflationary cosmology, we revise our initial descriptions about the early history of cosmic expansion in $\Lambda$CDM. Instead of a consistently slow expansion, we regard our large scale spatial universe as having undergone approximately exponential expansion for a period of time during our remote past, due (semiclassically) to the adiabatic vacuum effects of an `inflaton' field (or many such fields), before more familiar semiclassical physical processes take over (cf. footnote \ref{fnsemiclassicalinterpolation}). This rapid expansion washes out any interesting classical initial data in the early universe that could otherwise be relevant to the Cauchy problem specified at recombination in late-stage cosmology. 

Meanwhile, quantum fluctuations about the vacuum state of the inflaton are assumed to evolve semiclassically, comoving with the background cosmic expansion. This is how inflation may be regarded as a ``causal mechanism for generating'' (e.g. \citep[p. 177]{brandenberger2000inflationary}) the primordial perturbations that are ultimately responsible for seeding the evolution of non-trivial LSS in late-stage cosmology. For each instant in cosmic time during the inflationary period (i.e. the period of rapid cosmic expansion), we may think of the global vacuum physics of the inflaton in the following way. Across space about an observer in that instant, to linear order in perturbation theory, we may there decompose the global vacuum state of the inflaton into a homogeneous component (consistent with the background expansion) and an infinite collection of independent fluctuations on top, each indexed by its characteristic wavelength. Doing so, one finds that the sub-Hubble wavelength fluctuations proceed to evolve as independent harmonic oscillators, which comove--- and so, stretch--- with the rapidly expanding background off of that instant. Meanwhile, super-Hubble fluctuations resemble overdamped oscillators in the same. These latter modes--- whether they are associated with sufficiently low-frequency, infrared (IR) vacuum state physics in that instant or the stretched remnants of higher-frequency vacuum state physics at earlier instants--- thereby become physical candidates for classicalization in the quantum theory. And when they so classicalize, they thereby persist as frozen scalar degrees of freedom in spacetime at those same large scales, eventually to seed the evolution of LSS in late-stage cosmology.\footnote{What does classicalization look like in this proposal? It is a feature of the quantum field theory construction, given any strictly positive cosmic expansion in the background, that we may split super-Hubble modes into a quasi-isotropic term (i.e. a term which is compatible, on average, with the symmetries of the background expansion) and a noise term. The noise term may be identified as sub-Hubble, and so rapidly becomes negligible (provided that cosmic expansion is sufficiently rapid). But discarding the noise term altogether would amount to coupling the quasi-isotropic term to some trivial environment, in which case the quasi-isotropic term could be treated as a quantum system that has undergone decoherence \citep{polarski1996semiclassicality}. Taking this background-consistent decoherence as a stand-in for classicalization, that quasi-isotropic term propagates as an initially frozen mode according to a classical perturbation theory defined on the rapidly expanding background. For one dissent against such a decoherence view of the quantum-classical transition in inflation, see \citep{sudarsky2011shortcomings}.} Namely, following the end of the period of rapid expansion, they either come to seed background cosmic expansion in late-stage cosmology, or else they seed non-trivial LSS as they re-enter the Hubble radius and unfreeze prior to recombination (for more details, see \citep{brandenberger2004lectures}).

Note that, as just stated, this dynamics ties the duration of the period of rapid expansion in the early universe to an upper bound on the frequency of the quantum fluctuations that are associated downstream with the seeds of non-trivial LSS. This will become important in discussion of the TCC below. For any given length-scale of classical gravitational physics at recombination and any given frequency $\nu$ in the quantum theory relevant to the early universe, one may compute how long inflation \emph{would have had to last} there, such that the scalar degrees of freedom that seed, at recombination, non-trivial structure resolved at the given length-scale are sensitive to vacuum state physics whose origins were fluctuations above $\nu$ at a sufficiently early instant of cosmic time (and which then proceeded to stretch).

I will not discuss the details or merits of inflationary cosmology here. As has been stressed by \citet{brandenberger2011alternatives}, there exist alternatives to inflation that accomplish many of the same empirical feats. In what follows, this is precisely what I will take for granted, as pertains to any such discussion. There occurs, within the domain of early universe cosmology, some high energy physics responsible for producing all of what is observed in the CMB, thereby seeding the evolution of LSS in late-stage cosmology at recombination. This high energy physics admits of a description as a QFTCS construction (suitably interpreted), and is plausibly captured by a model of inflation. (Here, I set aside as too radical to countenance any alternatives to inflation that do not admit of descriptions as QFTCS constructions, for instance string gas cosmology. Such alternatives are models of entirely different theories of early universe cosmology, whose embrace by the community could dramatically change the role for trans-Planckian censorship in ongoing quantum gravity research, in ways that are difficult to trace in the abstract. See also remarks in footnote \ref{fnworkinghypothesis}.)

\subsection{Autonomy (or the necessary lack thereof)}

One might be surprised of the interplay, in the above description, between the semiclassical, high energy particle physics of structure formation and the classical, low energy evolution of LSS. In principle, one might even have liked it to be the case that the study of the evolution of LSS is autonomous with regards to any unknown or speculative high energy physics in our universe. I mean this in the sense discussed by \citep{batterman2018autonomy}, where, for instance ``our theories of `ordinary materials such as water, air, and wood' work remarkably well despite completely ignoring any structure of those materials at scales below centimeters'' \citep[note 3]{batterman2018autonomy}. So, in the present case, autonomy would mean that the study of the evolution of LSS could be pursued adequately by considering only those quantities deemed to exist at the largest, low energy scales, together with some known (and, presumably, classical) laws that we take to govern those quantities--- e.g. those of classical field theories.

Despite any such wish, it is clear that early universe cosmology is, as a sub-discipline within contemporary cosmology, predicated on a denial of such autonomy.\footnote{An important exception to this was Misner's program in chaotic cosmology, which was considered in the early days of the sub-discipline (cf. discussion by \citet[\S 5.5.1]{beisbart2009can} and references therein). Misner hoped that the high degree of spatial uniformity exhibited at recombination is an attractor in the classical dynamics of whatever theory ultimately governs the evolution of LSS. If his program had succeeded, the CMB could be considered, in a sense, generic with respect to initial conditions in our remote past, which were otherwise sensitive to high energy physics. In this way, the evolution of LSS could be decoupled from any high energy physics relevant within early universe cosmology, thereby recovering autonomy.} The point I would like to stress here, however, is that it is a denial of a very curious kind: minimal and controlled. In particular, the classical theory governing late-stage cosmology is, evidently, autonomous with regards to higher-energy, semiclassical characterizations of the `late-stage universe'. Likewise, the semiclassical theory governing early universe cosmology is, evidently, autonomous with regards to any lower-energy, classical characterization of the early universe. Where autonomy is denied is within the late-stage \emph{extension} of early universe cosmology, and also the early universe \emph{extension} of late-stage cosmology. And while we see no reason to take these extensions to be, in general, relevant to our narrative of cosmic history, we nonetheless demand that they agree about conditions that obtain along the shared boundary between them: recombination.\footnote{\citet{bursten2020function} argues that, in certain cases, conditions along a shared boundary between multiple models can play a `model-connecting function' in multiscale explanations. Recombination seems to fit this bill in the multiscale modeling of LSS in cosmology, even though the conditions that obtain there take the form of initial data (or final data, as it were) within each of the two relevant theories.}

To see what I mean by this, consider the manner by which inflationary cosmology rose to prominence as a program within early universe cosmology. Namely, the uniformity at recombination according to late-stage cosmology was regarded as fine-grained evidence of some distinctly high energy, dynamical process within the early universe, which subsequently gave rise to suitable conditions at recombination (and for which there is no relevant lower-energy physical description). In other words, the details observed within the CMB, as interpreted in the classical context of late-stage cosmology, were assumed, in the context of theory development in high energy particle physics, to be the right kind on which to rest an intricate, semiclassical account of structure formation within our remote past.\footnote{As widely popularized by \citet{guth1998inflationary}, the discovery of inflation was originally motivated by the prospects of using early universe cosmology specifically to study grand unified theories of particle physics (GUTs) in the framework provided by QFTCS. The `horizon problem' and `flatness problem' in the CMB data, referenced obliquely here, provided empirical constraints on the additional quantum fields whose presence was motivated by the viability of GUTs. While GUTs have fallen out of favor, inflation remains.} Conversely, descriptions given at high energy scales in early universe cosmology were here taken to matter to the Cauchy problem specified at recombination in late-stage cosmology. 

This perspective remains popular today. To wit: observations concerning the evolution of LSS in the late-stage universe provide an empirical window into high energy particle physics. In particular, this is because high energy particle physics in the early universe, autonomous with respect to any low energy physics there, is understood to seed the conditions that obtain at recombination. Those conditions are then taken to be relevant to the Cauchy problem in the particular low energy, autonomous classical theory that is taken to be descriptive of the evolution of LSS thereafter. 

\section{Introducing the Planck regime}

In the narrative rehearsed at the beginning of the previous section, early universe cosmology is born out of our reluctance to trust the relevance of a classical theory to the study of LSS, when ambient temperatures are sufficiently high. As just discussed, the converse is also true: we are simultaneously reluctant to trust the relevance of the high energy, semiclassical theory to the study of LSS, when ambient temperatures are sufficiently low.

But likewise, at least in the case of inflationary cosmology, temperatures eventually grow large enough--- as we move backward in cosmic time about any point in the early universe--- that we come to abandon trust in the semiclassical theory as well. Instead, we regard a hitherto-unknown theory of quantum gravity as that which reigns supreme. So defines the `Planck regime'--- a regime in which the bare structure of classical spacetime is discarded \citep{callender2001physics}. In this sense, the exit from the Planck regime gains the distinction of constituting the beginning of our large scale, (semi)classical universe.\footnote{\label{fnframe}Here, I take the exit from the Planck regime to be a term of art, which corresponds to an energy scale threshold above which we take ourselves to lack any means of quantifying our ignorance about the relevant physics (at least for as long as we lack a theory of quantum gravity). Note that setting up this correspondence is tantamount to specifying a frame on the relevant cosmological spacetime. But doing so is necessary, in order to regard quantum gravity as ``higher energy'' physics within our expanding universe. And, in any case, the construction is relevant to defining the notion of `scale' used to formulate effective field theories in the context of early universe cosmology, discussed below, despite the possibly rapid expansion of space in the background that is characteristic of inflationary cosmology.}

The Planck regime is sometimes referred to as the `Planck epoch' (e.g. in \citep{zinkernagel2008did}), but it is not properly conceived as situated temporally prior to cosmic expansion. This is a locus of persistent confusion in cosmological modeling: in fact, as will come up below, the exit from the Planck regime is perhaps better pictured--- at least, in a usual cosmological spacetime with always positive expansion--- as a spatially compact, timelike hypersurface, which is defined relative to any choice of inertial, stationary observer in the spacetime. The Planck regime, defined with respect to that observer, thereby refers to a well-defined region within the spacetime: in effect, a worldtube centered around that observer, whose spatial radius is fixed suitably near the initial singularity and is, thereon, decoupled from cosmic expansion (cf. footnote \ref{fnframe}). Within this region, we do not trust the theory to be descriptively accurate. Moreover, since cosmic expansion is always strictly positive, the normalized spatial volume of the region pinches off to zero over cosmic time (i.e. in coordinates comoving with expansion about the stationary observer's worldline). 

This asymptotic reasoning suggests one sense in which our ignorance about what goes on in the Planck regime is, with regards to LSS in our universe, plausibly of diminishing importance with cosmic time. This is particularly apt, in the context of interpreting the wholly classical physics used to model the evolution of LSS in late-stage cosmology. (By contrast, as will become clear, difficulties with employing this asymptotic reasoning in the early universe, at least in the case of inflationary cosmology understood as an effective field theory, are essential to contextualizing trans-Planckian censorship.)

As already discussed, inflation is immensely popular today as a particular inference concerning high energy particle physics, which came about by leveraging progress at the boundary shared by early universe and late-stage cosmology. One may thereby speculate that quantum gravity within the Planck regime may be constrained, similarly, by the expectation that such physics seeds the relevant conditions within early universe cosmology, at the exit from the Planck regime. In the particular case of inflationary cosmology, this is to inquire: how does the physics operating within the Planck regime give rise to the conditions associated with inflation in our remote past?\footnote{\label{fnhiccup}In other words, one may take the relevant model of inflation itself within our cosmic history as that which the Planck regime must output for early universe cosmology (see, e.g., \citep{easther2001inflation,martin2002cosmological,danielsson2006inflation}). Of course, there is a hiccup to overcome in this speculative analogy: as mentioned already, the exit from the Planck regime, considered as a surface in a cosmological spacetime, is not spacelike (as is true of recombination), but is instead timelike, spatially compact, and defined relative to an observer. Consequently, the exit from the Planck regime is not suitable as an initial data surface in classical or semiclassical field theories. Yet still, there is a sense in which quantum gravity relevant to modeling the Planck regime about an observer determines what I will below call ``pseudo- initial data'': physics that may be associated with such a surface, which the ensuing model of inflation is demanded to reproduce, along the same. Holographic approaches might be attractive here, which enforce a correspondence between some bulk region of a spacetime and its boundary.}

\subsection{The trans-Planckian problem(s) in cosmology}

I have just suggested that quantum gravity in the Planck regime constitutes a target of investigation that we hope to learn about, in virtue of the role it plays seeding conditions at the exit from the Planck regime within early universe cosmology. I have tried to paint this picture in such a way that, given the inference to a model of structure formation in the early universe, the study of the Planck regime is a natural follow-up. In both cases, the essential insight is that autonomy may be denied, in a minimal and controlled manner (i.e. along particular surfaces in the spacetime, serving as mutually constraining boundaries), to fruitful ends. Increasingly high energy physics matters to understanding the evolution of LSS observed today: in the first case, in virtue of structure formation at recombination; in the second case, in virtue of effecting the means of structure formation at the exit from the Planck regime. 

But there is at least one major disanalogy between the two cases, ostensibly owing to the timelike character of the exit from the Planck regime. Namely, the latter is plagued by the `trans-Planckian problem' in cosmology.\footnote{This is not to be confused with the trans-Planckian problem with Hawking radiation, which concerns the physical origins of certain thermal radiation that is otherwise associated with the global spacetime structure of black holes. For discussion of contemporary attitudes regarding this other problem, see \citep{wallace2018case} and references therein. Whether there are fruitful analogies to be drawn between the two subjects is worth philosophical scrutiny. In particular, the apparent success of the (timelike) membrane paradigm in defusing some of the objections to the physicality of Hawking radiation could suggest that the same framework may be extended, to defuse much of the setup discussed presently that motivates the TCC. On the other hand, see \citep{bedroya2021sitter} for some reservations about this argumentative tack.} As will become clear, reasoning from what I will soon dub the `weak' version of the problem makes it difficult to deny autonomy in early universe cosmology, as we typically understand it, with regards to quantum gravity in the Planck regime. And this is unfortunate. If autonomy cannot plausibly be denied, we lose a means of empirically constraining quantum gravity research based on cosmological evidence, which is otherwise secured by the relevance of quantum gravity to the Planck regime. Hence, the stage will be set for trans-Planckian censorship in the next section: the TCC stipulates that it is explicitly due to facts about quantum gravity, as become relevant in the Planck regime, that autonomy cannot plausibly be denied within early universe cosmology, at least as we typically understand it. Consequently, given an embrace of a suitable version of the TCC in one's candidate quantum gravity approach, one proceeds to empirically leverage our typical understanding of the early universe in their quantum gravity research, by new means. 

So, what is the trans-Planckian problem in cosmology? Begin with the commentary that there can (conceivably) exist classical physical degrees of freedom in descriptions of the late-stage universe, whose origins are quantum processes above any arbitrary high energy cutoff. Of particular interest here are processes that originate at energy scales high enough to be associated with the hitherto undeveloped theory of quantum gravity. These processes are of interest because, given our present understanding of fundamental physics, we currently have very little idea how even to model our own ignorance about them. 

Nonetheless, in the context of cosmology, these processes may be readily associated with the Planck regime in our universe, extending as far as one should like to track into the asymptotic future. Meanwhile, classical physical degrees of freedom in the late-stage universe whose origins lie in the Planck regime may thereby provide much desired windows into quantum gravity. This is because any such `trans-Planckian' physics ought to matter to our record of the evolution of LSS, on which we have a comparatively firm grasp. In other words, provided that we find a record of trans-Planckian physics in our universe, we may proceed to leverage our understanding of the evolution of LSS to hopefully learn about quantum gravity within the Planck regime. The trans-Planckian problem in cosmology may therefore be summarized as follows: how might we recognize the presence of trans-Planckian physics in our empirical descriptions of the evolution of LSS? 

As I have already forecast, I think it is helpful to distinguish two different versions of the trans-Planckian problem in cosmology--- `strong' and `weak'---, which it will be helpful below to have drawn apart from one another. What I have in mind as the strong version of the problem is a version that will threaten our usual, classical inferences in late-stage cosmology to the traces of early universe cosmology that are present at recombination. By contrast, the weak version of the problem, assuming away the strong one, will threaten (only) our semiclassical inferences about the early universe, based on those same traces present along the shared boundary between the two domains.

To draw these two versions apart, first recall that our theory of late-stage cosmology admits of a description in terms of a Cauchy problem specified at recombination. Assuming that our current theory of late-stage cosmology is descriptively accurate, it follows that any record of trans-Planckian physics in the evolution of LSS may be taken as (already) encoded in the initial conditions present at recombination. Of course, we might imagine ways by which our theory of late-stage cosmology could be descriptively accurate after recombination, despite there being trans-Planckian physics originating at moments in cosmic time well within our late-stage universe, i.e. the domain of late-stage cosmology. 

If this were the case, the familiar bulk dynamical equations (together with auxillary details, e.g. boundary conditions in the form of initial data at recombination and constraints on global structure) would happen to provide accurate descriptions, narrowly construed, of the evolution of LSS in our late-stage universe. But the relationship those equations would have in the bulk to any underlying physical description of the same in terms of quantum gravity would be markedly indirect or subtle, in comparison with the usual expectation. Namely, they would not stand in anything like a simple quantum-to-classical limit approximation, as understood in the context of particle physics, where one expects to recover the classical bulk dynamics of the evolution of LSS as a coarse-grained description of the outcome of sequestering to a microscopic `environment' all physical degrees of freedom where quantum field theoretic effects are thought to matter.\footnote{For an example of some of the new subtlety that is required in such an imagined scenario, see the discussion in \citep{smolin2017mond} that there may be various distinct `cosmological constant' dominated classical limit regimes of the dynamics of quantum gravity.} 

For one project in this skeptical tradition, consider the demonstration in \citep{mersini2001relic}. In their case, dark energy is identified as a phenomenological artifact of semiclassical fluctuations produced consistently (by fiat) through the late-stage universe within the Planck regime, which disperse highly non-linearly (and so, rapidly freeze out as classical modes capable of back-reacting on cosmic expansion). In a similar vein, in the case of entropic gravity presented by \citep{verlinde2017emergent}, long-range correlations between spacetime points can give rise to a superfluid-like phenomenology in the dark sector at late stages, which is relevant to descriptions of the evolution of LSS we get from $\Lambda$CDM. In both cases, trans-Planckian physics produced at late-stages is regarded as mimicking consequences of what are otherwise attributed to particular classical states of affairs that obtain at recombination. In other words, these cases imagine that the successes of our theory of late-stage cosmology are misleading us about our late-stage universe. In particular, we imagine being misled in our regarding the evolution of LSS as fully determined by the conditions that obtain at recombination. 

But meanwhile, an assumption to the contrary is essential as a premise to classically infer the conditions at recombination, based on empirical observations of the late-stage universe. Since those inferences are what then, in turn, render the same observations empirical constraints within early universe cosmology, there is some methodological tension between standard practice in cosmology, as discussed in section \ref{sectionCosmologyandHEP}, and what is being suggested by these imaginative exercises. 

The strong version of the trans-Planckian problem concerns this methodological tension. On one hand, we may (per the dictate of the trans-Planckian problem) contemplate how trans-Planckian physics could originate late in cosmic history--- despite appearances--- and thereby lose our primary means of constraining our high-energy theorizing about structure formation in the early universe. On the other hand, we may assume that the autonomy of late-stage cosmology, with regards to quantum gravity, is essentially correct, so as to secure the classical inference needed to constrain theorizing about the early universe.  But then, having assumed autonomy following recombination, we are just as well assuming that there is no trans-Planckian physics originating after recombination in our late-stage universe, which would otherwise help constrain quantum gravity research.\footnote{The statement of the methodological tension rehearsed here is likely too strong. There are caveats, but these should not be too important in the present discussion. For instance, certain phenomenological models of trans-Planckian physics in the dark sector in late-stage cosmology could turn out to have rather marginal effects on our thinking about recombination, at least as concerns the subsequent use of those effects (or their absence) to evidentially constrain inflationary dynamics in the early universe.}

In this case, the most substantive constraint on quantum gravity research that we would get from our empirical record of the late-stage universe, at least via late-stage cosmology alone, is a demand that the evolution of LSS turns out to stand in a suitable limit of the theory of quantum gravity sought. This is to demand that, once we have that theory of quantum gravity, there is, built into that theory, a means of articulating an in-principle reduction of late-stage cosmology: i.e. from the classical theory, to a theory of quantum cosmology. But, as \citet{batterman2018autonomy} points out in the case of continuum theories of bulk materials, the multiple realizability of any such autonomous, high-level continuum theory (in our case, the theory of the evolution of LSS) spoils any sense in which the in-principle reduction explains the success of the descriptions we get from that theory. This is because an ultimately adequate explanation of that success would rather consist in a demonstration that, once there is such an in-principle reduction provided (that is, once we have a low-level theory that recovers, where appropriate, the theory of late-stage cosmology), the descriptions obtained from the high level theory that so reduce are nonetheless robust under certain kinds of variations to the descriptions apt for the lower level--- i.e. those which, in principle, \emph{would give rise to} the empirically successful descriptions at the higher level. Alas, if the recovery of the high-level theory, i.e. the evolution of LSS, is not itself what explains the empirical successes of that high-level theory, then the successful empirical claims within that high-level theory cannot be regarded as the kind of empirical evidence to which the lower level theory is \emph{itself} beholden. 

The upshot is that, inasmuch as such a demand on the future theory of quantum gravity can be said to constrain research, it is a fairly weak demand: a multiscale consistency check on the quantum gravity theory applied to a given modeling context, rather than an empirical test of it. In this sense, the evolution of LSS is no more informative about quantum gravity in our universe than, say, a solid wooden table is informative of the same. This is the strong version of the trans-Planckian problem in cosmology: in order to leverage our typical understanding of the late-stage universe so as to (substantively) empirically constrain quantum gravity research, we lose our empirical window into the high-energy physics of the early universe. In other words, our capacity to mine empirical descriptions in late-stage cosmology for substantive empirical constraints on quantum gravity research is precluded by our standard understanding of structure formation in the early universe--- as that which is empirically constrained by its seeding the evolution of LSS at recombination.\footnote{Per the previous footnote, there are exceptions to this conclusion, concerning dark sector research. And still other empirical constraints on quantum gravity may be constructed out of our late-stage cosmological record, if one is happy to abandon or replace our understanding of structure formation in the early universe, which section \ref{sectionCosmologyandHEP} identifies as a part of standard practice.}

The second version of the trans-Planckian problem--- the `weak' problem--- begins with the ``less imaginative'' possibility as to what a record of trans-Planckian physics looks like in our late-stage universe. Namely, if some trans-Planckian physics were to originate within early universe cosmology, it could leave an imprint on the CMB at recombination, partially seeding the classically understood evolution of LSS in late-stage cosmology. If we set aside the strong problem just discussed (that is, assuming away the trans-Planckian problem for classical, relativistic continuum theory modeling in late-stage cosmology), we may push the conversation entirely back in time. For the evolution of LSS to offer a trans-Planckian window into quantum gravity, we ask: what does trans-Planckian physics look like in our empirical descriptions of the early universe, as provided by our current best theory of that domain? 

As discussed above, our theory of early universe cosmology takes the form of a semiclassical interpretation of QFTCS, where, at least in the case of inflationary cosmology, the relevant QFTCS construction additionally semiclassically sources cosmic expansion in the background. But in order to proceed in the present investigation, some greater detail will be necessary. In particular, these QFTCS constructions are supposed to wear their ultraviolet (UV) frequency cutoffs on their sleeves. In the context of an effective field theory, this can be formalized as a stipulation that all further degrees of freedom above that high energy, UV cutoff have been absorbed into coupling constants within the relevant construction (see, e.g., \citep{rivat2020philosophical}, or else the pedagogical presentation in \citep{polchinski1992effective}). But in the context of QFTCS, understood as an attempt to semiclassically approximate the effects of quantum gravity in the presence of quantum fields, we simply assume that there is some such cutoff scale.\footnote{We also assume that topological degrees of freedom in the classical theory of gravity only matter above that scale, so that we may freely make use of global spacetime properties in selecting a vacuum state relevant to the physics below the cutoff. Relatedly, cosmologists often proceed as if conditions along a spacelike initial data surface sufficiently far back in cosmic time determine such global structure properties of the QFTCS constructions that are thought relevant in cosmological modeling, immediately thereafter. But it seems to me that this is precisely to assume away the weak version of the trans-Planckian problem in cosmology, which concerns an exit from the Planck regime that spoils--- about the observer--- the sense in which the early chosen surface can be seen as Cauchy. This may yet be defensible; but, together with our having assumed away the strong version of the problem, it would seem at least to come at a discouraging procedural cost of weakening the evidential link between cosmological data and ongoing fundamental physics research.} The upshot is that we declare, somewhat artificially, a frequency mode above which we throw out any information within the QFTCS construction. 

This is the Planck regime within the construction: information that we do not trust to be descriptively accurate, on the basis that we lack a field theory of quantum gravity that happens to agree with such descriptions (and according to which we may otherwise explicitly recover the QFTCS construction as a low energy effective field theory). From this perspective, trans-Planckian modes are any low-energy artifacts of the UV regime within a QFTCS construction, which depend on features of the construction above the UV cutoff. We do not trust them to be accurate of low energy physics being modeled by the construction, precisely because we do not trust the UV regime to be accurate of high energy physics.

In \citep{brandenberger2013trans}, this state of affairs was formalized, at least in the context of cosmological spacetimes, in an elucidating way. First, one specifies a spatial origin point, from which the exit from the Planck regime is defined for all cosmic time in the early universe, as described above: as a spatially compact, timelike hypersurface centered on that origin point, with radius on the order of a Planck-length (in coordinates appropriate for labeling frequency modes in a perturbation theory at any point in cosmic time, independent of cosmic expansion).\footnote{As suggested in footnote \ref{fnframe}, identifying such a surface in the spacetime is tantamount to imposing a (global) frame in addition to that which defines stationarity (i.e. in addition to the frame that co-moves with cosmic expansion).} This is something like an initial data surface (to be sure--- a pseudo- initial data surface, cf. footnote \ref{fnhiccup}) that encodes all Planckian degrees of freedom in early universe cosmology, defined with respect to an inertial observer comoving with cosmic expansion. It also clearly delimits the Planck regime: that which, with respect to the choice of observer, brings about some pseudo- initial data on that surface (again, cf. footnote \ref{fnhiccup}).

Wherever the conditions along the exit from the Planck regime are well-described in terms of semiclassical fluctuations radiating off of it, it must be that those fluctuations either do not classicalize (in which case, they would not seed the evolution of LSS), or else they just so happen to agree with that dictated by the relevant low energy QFTCS construction, considered about an observer, which we have previously inferred about our remote past on the basis of considerations from late-stage cosmology. In other words, either there is no record of trans-Planckian physics at recombination, or else our usual model of structure formation is, by coincidence, predictive of a regime it was explicitly engineered so as not to predict. This is the weak problem: in order to leverage our typical understanding of the early universe so as to (substantively) empirically constrain quantum gravity research, we must assume that our model of the early universe is predictive of a regime it was not meant to predict. 

Taking this assumption to be clearly implausible--- an abuse of the usual formalism that we otherwise typically take to describe the early universe---, one concludes that there is no trans-Planckian physics in the early universe, which would otherwise come to seed the evolution of LSS at recombination. Early universe cosmology is, at least as we typically understand it, autonomous with regards to quantum gravity in the Planck regime. So, just as in the discussion above about late-stage cosmology, the only constraint on quantum gravity research provided by early universe cosmology is that of a multiscale consistency check. But then, having already set aside the possibility of trans-Planckian physics originating post-recombination, we are left with simply no means of learning about quantum gravity, via a trans-Planckian window in cosmology. Empirical claims from cosmology fail to constrain theory development in quantum gravity, in any substantive way. 

As the next section will emphasize, trans-Planckian censorship has the effect of turning this predicament on its head. An embrace of the TCC amounts to a declaration: the bare fact that denying autonomy in our typical high energy theorizing about structure formation is untenable is itself a substantive empirical constraint on our theorizing about quantum gravity. It is, per the embrace, due to facts about our universe within the Planck regime--- i.e. to be understood according to the future theory of quantum gravity--- that we may regard structure formation in the early universe as, inevitably, autonomous. In other words: yes, our descriptions of structure formation in the early universe would seem to absolutely lack trans-Planckian physics. But, per the embrace of the TCC, that there is such a lack is precisely what needs be addressed in the development of any theory of quantum gravity which, in light of our corpus of cosmological evidence, is to count as empirically adequate.

\section{The TCC}\label{secTCC}

I have, at last, reached the point where I may introduce the TCC, as it was presented in its original context. Here is the authors' own general description of the conjecture, in terms only slightly more technical than those of the slogan provided in the Introduction \citep[p. 4]{bedroya2019trans}:\footnote{\label{fnrenormalizability}The general framing here previews the view adopted in \citep{bedroya2021sitter}, wherein the TCC is suggested to constitute a common organizing principle behind various swampland conjectures in string theory [p. 6-7]: \begin{quote}``We  argue  that  TCC  could  be  viewed  as  a  natural  modification  of  the  renormalizability condition for gravitational theories where the conventional notion of renormalizability does not apply [...] Similar to how quantum field theory's consistency imposes renormalizability, it is natural to expect a UV-complete quantum theory of gravity must satisfy some renormalizability-like condition [...] In a sense, TCC is a natural gravitational analogue of the renormalizability condition in field theory.''\end{quote}.}
\begin{quotation} We conjecture that a field theory consistent with a quantum theory of gravity does not lead to a cosmological expansion where any perturbation with length scale greater than the Hubble radius trace [\emph{sic}] back to trans-Planckian scales at an earlier time.\end{quotation}
Putting aside, for the moment, what it means for a field theory to be consistent with a theory of quantum gravity, the spirit of the TCC is clear. Recall that fluctuations with length scale greater than the Hubble radius are what, in the context of inflationary cosmology, classicalize as initially frozen perturbations, with downstream consequences readily observable in the CMB; these are what seed the evolution of LSS in late-stage cosmology. Meanwhile, arbitrarily high energy modes which are associated with the global vacuum state of the field at arbitrarily early moments of the epoch of field-driven rapid expansion may be understood to dynamically stretch through that epoch. With enough time, these arbitrarily high energy modes may thereby stretch to super-Hubble scales, with all of the ensuing downstream consequences. 

The TCC therefore rules out, by fiat, the possibility that any quantum fluctuations that stretch and classicalize in this way encode the effects of quantum gravitational physics deep within our remote past. QFTCS constructions that do otherwise--- namely, permitting modes emanating from the exit from the Planck regime to grow larger than the Hubble radius in the course of rapid expansion--- are simply inconsistent with the future theory of quantum gravity sought. In particular, there is some feature of the quantum gravitational physics present in the Planck regime that spoils how long such a period of rapid expansion could last (at least, in any viable description of the early universe as such a QFTCS construction that sources rapid cosmic expansion). Nature precludes it.

In this way, the weak version of the trans-Planckian problem is circumvented by fiat: supposing the TCC, trans-Planckian physics in the early universe fails to become observable content within late-stage cosmology. As such, the TCC is manifestly a conjecture that our typical understanding of structure formation in the early universe is autonomous with regards to the quantum gravitational physics thought to govern the Planck regime. What seeds late-stage cosmology is (merely) high energy physics below the Planckian cutoff.

But crucially, the TCC states that it is \emph{because of quantum gravity} that this is so, rather than by cosmic accident. QFTCS dynamics like that ascribed to the inflaton in the early universe \emph{are prohibited from lasting long enough to do otherwise}, given the relevance of the future theory of quantum gravity to that domain. In this way, an embrace of the TCC is very much like what \citet[\S 11.3]{currie2018rock} evidently has in mind (see also \citep{currie2021science}), when he introduces the notion of an `empirically grounded speculation' in discussion of the epistemic role for pragmatism in the historical sciences. These speculations are scientific claims that are ``justified on their fruits'' [p. 288], in the course of ongoing research. In the present case, the speculation is that whatever determines the question of `consistency' of an arbitrary QFTCS construction in the theory of quantum gravity is also the reason that, due to the relevance of quantum gravity within the Planck regime in the early universe, there is ultimately no trans-Planckian physics present at recombination (at least, given our typical understanding of the early universe). The upshot is that, given a commitment to inflationary techniques modeling structure formation in early universe cosmology, empirical descriptions of structure formation in that domain are rendered effects of (UV) quantum gravity phenomenology (a provocative statement, to be sure--- see footnote \ref{fnswamp} below for qualifications).

This, of course, takes for granted that it makes sense to talk about QFTCS constructions being consistent or inconsistent with a theory of quantum gravity, absent recourse to the latter theory in advance of its own development. Or, to put the matter more strongly: the TCC takes for granted that we know what it means, in quantum gravity research, to conclude that the relevance of some QFTCS constructions at low energies may be discarded, in virtue of facts that will be identified about higher energy physics. Meanwhile, other such constructions persist as descriptive of possible physics, in virtue of those same future facts. The point here is that consistency must mean something other than logical consistency, where (for instance) any description of a quantum gravitational system in the terms of QFTCS fails. But nor can it mean something as weak as approximation: some possibilities in the limit need to be ruled out, by the very means available to articulate the appropriate limit, in the first place. 

Elsewhere \citep{schneider2020s}, I have argued that making assumptions about the future theory of quantum gravity sought is an ordinary state of affairs in the course of developing it. Following suit, it can be instructive to consider how much of the contour of the future theory is already decided, when one proceeds to develop the future theory on the basis of such assumptions. In the present case, the original authors of the TCC favor a string theory approach to quantum gravity. But their own general framing of the conjecture should make us inclined to think that not all of the toolkit of string theory is being used. As I will now claim, one needs only something in one's candidate quantum gravity approach that is analogous to the tools, known in string theory, as `swampland conjectures'. If there are such analogues, one may make use of the TCC in developing a theory of quantum gravity accordingly. 

In string theory, it is expected that for many QFTCS constructions with a stipulated UV cutoff, there exist string theoretic models that agree about all physics below that cutoff,  and meanwhile are UV-complete. In other words, the latter are descriptive at all energies, but their descriptions at low energies are identical to those provided by the former. The QFTCS constructions that are arranged in this correspondence form the low energy `landscape' of string theory. This landscape circumscribes all that is string theoretically possible, provided that our attention is restricted to the effects of that string theoretic physics that are below some or other high energy cutoff. Thus, within string theory research, the landscape constrains the modeling of low-energy empirical regimes by means of QFTCS constructions, understood as (correspondingly low-energy) effective field theories. 

As was first discussed in 2005, it need not be the case that all such QFTCS constructions reside in the landscape. Some reside in the `swampland', failing to admit any such UV-completion \citep{vafa2005string}. (Hence, the constraint discussed above, in the context of string theory research, is non-trivial.) Which constructions reside in the swampland is a matter investigated via `swampland conjectures': formal conjectures about families of such constructions that plausibly fail to admit UV-completions in the framework of string theory, because of obstructions that tend to arise in the attempt to explicitly construct the completions. As such, any particular swampland conjecture states that there is something distinctive of the formal apparatus of string theory that prohibits the descriptive relevance of any construction in the relevant family (provided that our universe is, ultimately, stringy).\footnote{\label{fnswamp}For a comprehensive review of the swampland, see \citep{palti2019swampland}. Above, I suggested that the TCC renders inflation in the early universe as effects of (UV) quantum gravity phenomenology. As the present discussion highlights, this is a consequence of there being architecture like the swampland in one's quantum gravity approach. But just as well, the differences between quantum gravity approaches will cast the same empirical descriptions of structure formation in the early universe as different \emph{kinds} of quantum gravity phenomenology. Depending on how one's quantum gravity approach organizes questions about the theoretical consistency of low-energy effective models, the early universe may empirically constrain theorizing in essentially different ways. So, for instance, many of the relevant technical obstructions in string theory concern moduli stabilization about a de Sitter vacuum state, given (per the TCC) the inflationary early universe as an empirical constraint on the program. But other quantum gravity approaches may run up against entirely different technical obstructions, given all the same.}

One such swampland conjecture provides the intellectual origins of the TCC: the family of constructions relevant to inflationary cosmology, which in particular raise the specter of the weak version of the trans-Planckian problem in cosmology, plausibly fails to admit stringy UV-completions. In \citep{saito2020trans}, it is challenged whether there is such a swampland conjecture, properly construed, that realizes the TCC (and which is consistent with current empirical commitments). Nonetheless, for present purposes, it is merely relevant that there exists such a tool as that provided by swampland conjectures in the relevant candidate quantum gravity approach, in order to formulate a version of the TCC. This is, after all, just what is needed to stipulate \emph{that} the TCC be rendered true of the future theory, as a condition of adequacy in the course of that theory's development in string theory. Similarly, in order for the TCC to constrain the development of the future theory in any candidate quantum gravity approach, an analogue to such a tool is needed in that approach. That is, we need a way of adjudicating which QFTCS constructions \emph{cannot be descriptive} of the low energy physics of the quantum gravity theory sought, via that approach. 

\section{Conclusion}

We have seen a little of how trans-Planckian censorship can help empirically constrain quantum gravity research, provided that one ensures that the relevant conditions of theorizing are met within their chosen candidate quantum gravity approach. The TCC amounts to an empirically grounded speculation about quantum gravity within the Planck regime, as is relevant to effecting the means of structure formation--- as we typically understand the subject--- in early universe cosmology. In this respect, embracing the TCC has the consequence of rendering inflationary descriptions of structure formation in the early universe as, in themselves, claims about quantum gravity phenomenology--- albeit in a manner that will, inevitably, be specific to one's favored quantum gravity approach. Namely: as a theory of quantum gravity is further developed within that approach, some or other swampland-like conjecture therein, which screens out trans-Planckian physics in the context of early universe inflationary dynamics, \emph{must come to be true}--- that is, lest the theory turn out to fail to account for all of the relevant empirical data.\footnote{\label{fnendorsement} Not addressed in this article is what it ultimately means for a theorist to have embraced the TCC (i.e. such as would supply the normative force behind the italicized portion of the final sentence in the article). In particular, it seems a tall order to claim that the pragmatic epistemic upshots discussed here are suitable as warrant for \emph{belief} in the TCC, and, indeed, it is in this sense that the embrace of the TCC rather amounts to speculation. (After all: it may be that there simply is no trans-Planckian physics in the early universe, in which case a theorist is certainly not compelled to believe that the absence of an effect there is due to some further, specific fact about quantum gravity.) Recent work on `rational endorsement' by \citet{fleisher2018rational} (see also \citep{fleisher2019endorsement}) might be helpful here, as a doxastic attitude that is distinct from belief, and which is more conducive to navigating the collective epistemic benefits of sustained peer disagreement in science.}

\bibliographystyle{abbrvnat}
\bibliography{transplanckianCensorship}
\end{document}